\documentclass[pra,twocolumn]{revtex4-2}

\usepackage[hidelinks]{hyperref}
\usepackage{amssymb}
\usepackage{array}
\usepackage{graphicx}
\usepackage{dsfont}
\usepackage{xcolor}

\definecolor{fulleq}{HTML}{4B0082}
\definecolor{bigeq}{HTML}{DAA520}
\definecolor{mueq}{HTML}{32CD32}
\definecolor{simpleq}{HTML}{4169E1}
\definecolor{mc}{HTML}{DC143C}

\begin{document}

\title{A simplified approach to the repulsive Bose gas from low to high densities and its numerical accuracy}

\author{Eric A. Carlen}
\affiliation{\it Department of Mathematics, Rutgers University}
\email{carlen@rutgers.edu}

\author{Markus Holzmann}
\affiliation{Univ. Grenoble Alpes, CNRS, LPMMC, 38000 Grenoble, France}
\affiliation{Institut Laue Langevin, BP 156, F-38042 Grenoble Cedex 9, France}
\email{markus.holzmann@grenoble.cnrs.fr}

\author{Ian Jauslin}
\affiliation{Department of Physics, Princeton University}
\email{ijauslin@princeton.edu}

\author{Elliott H. Lieb}
\affiliation{Departments of Mathematics and Physics, Princeton University}
\email{lieb@princeton.edu}

\begin{abstract}
In 1963, a {\it Simplified Approach} was developed to study the ground state energy of an interacting Bose gas with a purely repulsive potential.
It consists in the derivation of an Equation, which is not based on perturbation theory, and which gives the exact expansion of the energy at low densities.
This Equation is expressed directly in the thermodynamic limit, and only involves functions of $3$ variables, rather than $3N$.
Here, we revisit this approach, introduce two more equations and show that these yields accurate predictions for various observables for {\it all} densities for repulsive potentials with positive Fourier transform.
Specifically, in addition to the ground state energy, we have shown that the Simplified Approach gives predictions for the condensate fraction, two-point correlation function, and momentum distribution.
We have carried out a variety of tests by comparing the predictions of the Equations with Quantum Monte Carlo calculations for exponential interaction potentials as well as a different, finite range potential of positive type, and have found remarkable agreement.
We thus show that the Simplified Approach provides a new theoretical tool to understand the behavior of the many-body Bose gas, not only in the small and large density ranges, which have been studied before, but also in the range of intermediate density, for which much less is known.
\end{abstract}

\maketitle

\section{Introduction}\label{sec:intro}
\indent
Bose gases are one of the foundational objects in the statistical mechanics of quantum systems, and have been the focus of much scrutiny, dating back to the early days of quantum mechanics\-~\cite{Le29}.
Nevertheless, there are still several important problems to be solved, in the case of interacting Bose gases, in which the correlations between particles make the analysis very difficult.
In this case, observables may be computed by either performing numerical computations using finite-size approximations and extrapolations, or by devising effective theories which capture some of the correlations between particles, while remaining integrable.
In this paper, we present an effective theory which goes back to 1963\-~\cite{Li63}, and which we have found gives accurate predictions in the thermodynamic limit at {\it all} densities that have been verified numerically by Quantum Monte Carlo (QMC) computations.
This remarkable agreement leads us to suggest that this may be a new way of understanding and analyzing the quantum many-body problem.
\bigskip

\indent
In the low density regime, an effective theory which has proved to be extremely successful is due to Bogolubov\-~\cite{Bo47}, who devised a scheme in which the many body-Hamiltonian is reduced to a quadratic operator, which captures pair correlations rather well, and, at the same time, can be explicitly diagonalized (see\-~\cite{ZB01} for a review).
By applying Bogolubov's scheme to an idealized Hamiltonian in which the interaction potential $v$ is replaced by a localized {\it pseudo-potential}, Lee, Huang and Yang derived a large collection of predictions for the Bose gas at low density.
In particular, they computed that the ground state energy per-particle should behave as\-~\cite[(25)]{LHY57}:
\begin{equation}
  e_0=2\pi\rho a_0\left(1+\frac{128}{15\sqrt\pi}\sqrt{\rho a_0^3}\right)
  \label{lhy}
\end{equation}
where $\rho$ is the particle density, $a_0$ is the scattering length of $v$ (throughout this paper, we will take $\hbar=m=1$).
The leading order term $2\pi\rho a_0$ is originally due to Lenz\-~\cite{Le29}.
The Lee-Huang-Yang formula\-~(\ref{lhy}) can also be derived from the computation done by Bogolubov\-~\cite{Bo47,Li65}.
This expansion is {\it universal}, in that it only depends on the scattering length $a_0$, and not on the details of the potential.
Lee, Huang and Yang also made a prediction for the ground state non-condensed fraction $\eta_0$, that is, the fraction of particles that are {\it not} in the Bose-Einstein condensate\-~\cite[(41)]{LHY57}:
\begin{equation}
  \eta_0=\frac{8\sqrt{\rho a_0^3}}{3\sqrt\pi}
  .
  \label{eta0}
\end{equation}

\indent
After much work over more than sixty years, it was finally proved\-~\cite{Dy57,LY98,ESY08,YY09,GS09,BBe19,BS20,FS20} that\-~(\ref{lhy}) is asymptotically correct at low densities.
The formula for the non-condensed fraction\-~(\ref{eta0}) has, to this day, not been proved to hold for the interacting Bose gas in the thermodynamic limit, though it has been confirmed by numerical experiments\-~\cite{GBC99}.
\bigskip

\indent
Concerning the  ground state energy at high densities, it has been shown\-~\cite{Li63} that if the potential is of positive type (non-negative with a non-negative Fourier transform), then, as $\rho\to\infty$,
\begin{equation}
  e_0\sim\frac\rho2\int d\mathbf x\ v(\mathbf x)
  .
  \label{ehigh}
\end{equation}
The positivity of the Fourier transform of the potential is required for this to hold.
In fact, S\"ut\H o\-~\cite{Su11} has proved that, for the classical Bose gas (at asymptotically large densities, for many potentials, the classical ground state coincides with the quantum one), the high-density ground state is uniform for positive type potentials, but it exhibits periodic patterns for certain potentials that are not of positive type.
In the latter case, (\ref{ehigh}) cannot possibly hold.
In Section\-~\ref{sec:limits}, we will discuss a simple example of a potential that is not of positive type for which $e_0/\rho\to0$.
From now on, we will restrict our attention to potentials of positive type.
The asymptotic formula\-~(\ref{ehigh}) coincides with the ground state energy in Hartree theory, in which all Bosons are assumed to be condensed.
Note that, whereas Hartree theory is accurate at asymptotically large densities, there are various effective theories that produce accurate results for large finite densities, such as those based on the Random Phase Approximation and the Mean Spherical Approximation (MSA)\-~\cite{Kr02}.
\bigskip

\indent
Therefore, the Bose gas is described by Bogolubov theory at low density, and Hartree theory or the MSA at high density.
In this paper, we will discuss another effective theory for the ground state of the repulsive Bose gas with a positive type potential, which is highly accurate at all densities, which is {\it exact} at low and high densities, and highly accurate at all intermediate densities.
In other words, it is a physically descriptive interpolation between Bogolubov and Hartree theory.
To justify our claim that it is in good {\it quantitative} agreement with the physics at all densities, we rely on with QMC simulations of the Bose gas for intermediate densities.
This equation was originally introduced in 1963\-~\cite{Li63}, and studied for the high density Jellium\-~\cite{LS64}, and in one dimension\-~\cite{LL64}.
There has been no research progress since then.
The merit of this equation is twofold.
First, it provides a tool to study the Bose gas at intermediate densities, about which little is known, and, since the Bose gas is strongly correlated in this regime, we expect the physical behavior of the system to be significantly different from the low and high density limits.
Second, the approach leading to this equation is quite different from Bogolubov theory, so it may shine a new light on the low density physics of the system, and, perhaps, lead to progress in the proof of the existence of Bose-Einstein condensates at small positive densities.
\bigskip

\indent
The effective theory described in this paper gives a prediction for a function derived from the ground state wave-function $\psi_0$ of the Bose gas in the thermodynamic limit, which is automatically symmetric and non-negative:
\begin{equation}
  g_2(\mathbf x_1-\mathbf x_2):=\lim_{\displaystyle\mathop{\scriptstyle N,V\to\infty}_{\frac NV=\rho}}\frac{\int \frac{d\mathbf x_3}{V}\cdots \frac{d\mathbf x_N}{V}\ \psi_0(\mathbf x_1,\mathbf x_2,\cdots,\mathbf x_N)}{\int \frac{d\mathbf y_1}{V}\cdots \frac{d\mathbf y_N}{V}\ \psi_0(\mathbf y_1,\cdots,\mathbf y_N)}
  .
  \label{g2}
\end{equation}
The function $g_2$ can be interpreted as the two-point correlation function of the probability distribution $\psi_0\geqslant 0$ (suitably normalized).
Note that this is different from the quantum probability distribution $|\psi_0|^2$.
The effective theory gives a prediction, denoted by $u$, for an approximation of $1-g_2(\mathbf x-\mathbf y)$.
This prediction satisfies the following equation\-~\cite{Li63}
\begin{equation}
  (-\Delta+v(\mathbf x))u(\mathbf x)
  =v(\mathbf x)
  -\rho(1-u(\mathbf x))(2K(\mathbf x)-\rho L(\mathbf x))
  \label{fulleq}
\end{equation}
with
\begin{equation}
  K(\mathbf x):=\int d\mathbf y\ u(\mathbf y-\mathbf x)S(\mathbf y)\equiv u\ast S(\mathbf x)
  \label{K}
\end{equation}
\begin{equation}
  S(\mathbf x):=(1-u(\mathbf x))v(\mathbf x)
  \label{S}
\end{equation}
\begin{equation}
  \begin{array}{>\displaystyle l}
    L(\mathbf x):=
    \int d\mathbf yd\mathbf z\ u(\mathbf y)u(\mathbf z-\mathbf x)
    \cdot\\\cdot
    \left(
      1-u(\mathbf z)-u(\mathbf y-\mathbf x)+\frac12u(\mathbf z)u(\mathbf y-\mathbf x)
    \right)S(\mathbf z-\mathbf y)
    .
  \end{array}
  \label{L}
\end{equation}
This equation will be called the {\bf Full Equation}, as we will also be considering a hierarchy of three approximations to this equation:
\begin{itemize}
  \item the {\bf Big Equation} (which will be rendered in plots in {\bf\color{bigeq} yellow}), in which we neglect the $\frac12u(\mathbf z)u(\mathbf y-\mathbf x)$ term in\-~(\ref{L}):
  \begin{equation}
    -\Delta u(\mathbf x)
    =
    (1-u(\mathbf x))\left(v(\mathbf x)-2\rho K(\mathbf x)+\rho^2 L_{\mathrm{bigeq}}(\mathbf x)\right)
    \label{bigeq}
  \end{equation}
  with
  \begin{equation}
    L_{\mathrm{bigeq}}:=
    u\ast u\ast S
    -2u\ast(u(u\ast S))
    .
    \label{Lbigeq}
  \end{equation}

  \item the {\bf Medium Equation} ({\bf\color{mueq}green}), in which we further neglect the $2u\ast(u(u\ast S))$ term in\-~(\ref{Lbigeq}), and drop the $u(\mathbf x)$ in the $(1-u(\mathbf x))$ prefactor of $K$ and $L_{\mathrm{bigeq}}$ in\-~(\ref{bigeq}):
  \begin{equation}
    -\Delta u(\mathbf x)
    =
    (1-u(\mathbf x))v(\mathbf x)-2\rho K(\mathbf x)+\rho^2L_{\mathrm{mueq}}(\mathbf x)
    \label{mueq}
  \end{equation}
  with
  \begin{equation}
    L_{\mathrm{mueq}}:=u\ast u\ast S
    .
    \label{Lmueq}
  \end{equation}

  \item the {\bf Simple Equation} ({\bf\color{simpleq}blue}), in which we further approximate $S$ by $\delta(\mathbf x)\frac{2\tilde e}\rho$ in\-~(\ref{K}) and\-~(\ref{Lmueq}):
  \begin{equation}
    (-\Delta+v(\mathbf x) + 4\tilde e)u(\mathbf x)
    =v(\mathbf x)
    +2\tilde e\rho u\ast u(\mathbf x)
    \label{simpleq}
  \end{equation}
  with
  \begin{equation}
    \tilde e=\frac\rho2\int d\mathbf x\ (1-u(\mathbf x))v(\mathbf x)
    .
  \end{equation}

\end{itemize}
The basis for making these approximations is discussed in section\-~\ref{sec:approx}.
The Big Equation is easier to solve numerically than the Full Equation, yet it remains very accurate.
However, the mathematical analysis of the Full, Big  and Medium Equations is quite difficult and so far has not been accomplished. 
In this regard, the situation is much better for the Simple Equation, for which a well-developed  mathematical study has been carried out in \cite{CJL20,CJL20b}, and it is also quite simple to investigate its solutions numerically.
The Medium Equation also has this latter advantage; it has a simpler structure than the Big Equation and is considerably easier to solve numerically.  
As we show here it gives good results over a wider range of densities than the Simple Equation.
\bigskip

\indent
The Simple Equation nonetheless gives accurate results at least for low and high densities, for which it yields asymptotically correct results.
In a previous publication\-~\cite{CJL20}, we proved that the Simple Equation predicts an energy that coincides asymptotically with\-~(\ref{lhy}) at low density, and with\-~(\ref{ehigh}) at high density.
In another paper\-~\cite{CJL20b}, released concurrently with the present paper, we prove that the condensate fraction predicted by the Simple Equation agrees asymptotically with\-~(\ref{eta0}) at low density.

\indent
In the present paper, we discuss some more quantitative results, with more of a focus on the Big Equation, which we have found to be very accurate by comparing its predictions to Quantum Monte Carlo simulations.
We will consider potentials that are of positive type, with a special focus on exponential potentials of the form $\alpha e^{-|\mathbf x|}$.
We have found that the prediction for the energy is very accurate for {\it all} densities, see Figure\-~\ref{fig:energy}.
In the case $\alpha=1$, the relative error compared to the QMC simulation is as small as $0.1\%$, and is comparable to the error made by a Bijl-Dingle-Jastrow function Ansatz \cite{Bi40,Di49,Ja55}, see Figure\-~\ref{fig:cmp_energy}, even though the solution of the Big Equation is much easier to compute numerically than the Bijl-Dingle-Jastrow optimizer.
The prediction for the condensate fraction is less accurate in the intermediate density regime, though still remarkably good for small values of $\alpha$, see Figure\-~\ref{fig:condensate0.5}.
For larger $\alpha$, the Big Equation is off the mark, see Figure\-~\ref{fig:condensate16}, although the qualitative features of the condensate fraction are still well reproduced.
We have also carried out similar computations for the hard core potential, for which we also find good agreement, see Figure\-~\ref{fig:hardcore}.

\indent
Because computing with the Big Equation is relatively easy from a computational point of view, we have been able to probe some observables in the intermediate density regime, far from the low density Bogolubov regime and the high density mean field regime.
Comparing to QMC simulations, we have found that $g_2$ (see\-~(\ref{g2})) is accurately predicted by both the simple and the Big Equations at low density, but, as the density is increased, the prediction from the Simple Equation drops away abruptly, but the Big Equation remains accurate: see Figure\-~\ref{fig:ux}.
When this occurs, a maximum that is $>1$ appears, thus indicating that there is a new length scale appearing in the problem, at which there is a small increase in the probability of finding a particle.
This picture also holds for the usual quantum two-point correlation function, which we can also predict rather accurately, see Figure\-~\ref{fig:correlation}.
This suggests a non-trivial, strongly coupled phase at intermediate densities, which was thus predicted by the Big Equation, and validated by QMC simulations.
\bigskip

\indent
Naturally, this is not the first investigation into strongly coupled Bose gases.
Indeed, there has been much interest lately in the {\it unitary Bose gas}, in which case the interaction potential is a Dirac delta function (a contact interaction), and the scattering length is taken to infinity (see\-~\cite{CGe10} for a review).
Increasing the scattering length results in non-trivial many-particle effects, such as the appearance of Efimov trimers\-~\cite{Ef70,KMe06,NE17}.
This can be seen\-~\cite{CW11,MKe14,SBe14,KXe17,FLe17} in terms of the {\it universal} Tan relation\-~\cite{Ta08a}, which states that the momentum distribution $\mathcal M(\mathbf k)$ satisfies, at large $\mathbf k$,
\begin{equation}
  \mathcal M_0(\mathbf k)\sim\frac{c_2}{|\mathbf k|^4}
  ,\quad
  c_2=8\pi a_0^2\frac{\partial e_0}{\partial a_0}
  .
  \label{tan}
\end{equation}
For the Big and Simple Equations discussed in this paper, we have found that this relation holds in the range
\begin{equation}
  \sqrt{\rho a_0}\ll|\mathbf k|\ll1
\end{equation}
which is another confirmation of the accuracy of the effective equation at small densities.
However, if $\sqrt\rho\gtrsim1$, then the universal Tan regime does not exist, and the picture in terms of strongly coupled few-particle configurations inherent to the analysis of unitary Bose gases\-~\cite{CW11,SBe14} breaks down, as the Bose gas transitions to a strongly correlated liquid.
This is confirmed for the prediction of the Big Equation, see Figure\-~\ref{fig:tan}.
\bigskip

\indent
As further evidence of the breakdown of universality in the intermediate density regime, we have also compared the ground state energy for two very different potentials, which have the same scattering length and the same integral.
We have found that the energy for these two potentials is significantly different in the intermediate density regime, see Figure\-~\ref{fig:compare_pots}.
For these two potentials, we have also found that the Quantum Monte Carlo data fits very well with the prediction of the Big Equation.
\bigskip

\indent
The rest of the paper is structured as follows.
In section\-~\ref{sec:approx}, we detail the approximation needed to get from the many-body Bose gas to the Full Equation, and then discuss the approximations leading to the Big, Medium and Simple Equations.
In section\-~\ref{sec:montecarlo}, we compare various physical quantities predicted by these equations to QMC simulations of the Bose gas.
In section\-~\ref{sec:hardcore}, we treat the hard core potential.
In section\-~\ref{sec:limits}, we discuss the limitations of the approximations.

\section{Derivation of the Full Equation and its approximations}\label{sec:approx}
\indent
Let us now discuss the derivation of the Full Equation, which follows\-~\cite{Li63}, and the approximations that lead to the Big, Medium and Simple Equations.
Whereas this derivation is based on uncontrolled approximations, it is justified by the remarkable accuracy of the resulting predictions compared to QMC computations.
We start from the many-body Hamiltonian: denoting the number of particles by $N$,
\begin{equation}
  H=-\frac12\sum_{i=1}^N\Delta_i+\sum_{1\leqslant i<j\leqslant N}v(\mathbf x_i-\mathbf x_j)
\end{equation}
(we set $\hbar=m=1$).
We confine the $N$ particles in a cubic box $\Lambda$  of volume $V$, and impose periodic boundary conditions.
Later on, we will take the thermodynamic limit $N,V\to\infty$, $\frac NV=\rho$ fixed.

\indent
In the derivation presented here, we will rely on the translation invariance of the Hamiltonian, which does not allow us to study a system with a trapping potential at this time.

\indent Let  $E_N$ denote the ground state energy and let $\psi_N(\mathbf x_1,\cdots,\mathbf x_N)$  denote the ground state wave function so that
\begin{equation}
  H\psi_0(\mathbf x_1,\cdots,\mathbf x_N)=E_N\psi_N(\mathbf x_1,\cdots,\mathbf x_N)
  \label{eigval}
\end{equation}
where $v\geqslant0$ is an integrable pair potential.
Instead of taking the scalar product of both sides of the equation with $\psi_0$, which would yield an expression relating the ground state energy to the 1-particle reduced density matrix, we will simply integrate both sides of the equation, and find that, using the translation invariance of the system,
\begin{equation}
  \frac{E_N}N=\frac{N-1}{2V}\int d\mathbf x\ v(\mathbf x)g_2^{(N)}(\mathbf x)
  \label{EN}
\end{equation}
with
\begin{equation}
  \begin{array}{>\displaystyle l}
    g_n^{(N)}(\mathbf x_2-\mathbf x_1,\cdots,\mathbf x_N-\mathbf x_1):=
    \\[0.3cm]\hskip20pt:=\frac{\int\frac{d\mathbf x_{n+1}}V\cdots\frac{d\mathbf x_N}V\ \psi_0(\mathbf x_1,\cdots,\mathbf x_N)}{\int\frac{d\mathbf x_{1}}V\cdots\frac{d\mathbf x_N}V\ \psi_0(\mathbf x_1,\cdots,\mathbf x_N)}
    .
  \end{array}
  \label{g}
\end{equation}
In particular, note that the kinetic term has disappeared entirely.
Furthermore, by the Perron-Frobenius theorem, $\psi_0\geqslant 0$, so $g_n^{(N)}$ can be interpreted as the $n$-point correlation function of the probability distribution $\psi_0$ (suitably normalized) which is not the usual quantum probability distribution.

\indent
We can then express $g_2^{(N)}$ by integrating\-~(\ref{eigval}) with respect to $\mathbf x_3,\cdots,\mathbf x_N$: using the translation invariance of the system,
\begin{widetext}
\begin{equation}
  \begin{array}{>\displaystyle l}
    -\Delta g_2^{(N)}(\mathbf x-\mathbf y)
    +v(\mathbf x-\mathbf y)g_2^{(N)}(\mathbf x-\mathbf y)
    +\frac{N-2}V\int d\mathbf z\ (v(\mathbf x-\mathbf z)+v(\mathbf y-\mathbf z))g_3^{(N)}(\mathbf y-\mathbf x,\mathbf z-\mathbf x)
    \\[0.5cm]\hfill
    +\frac{(N-2)(N-3)}{2V^2}\int d\mathbf zdt\ v(\mathbf z-t)g_4^{(N)}(\mathbf y-\mathbf x,\mathbf z-\mathbf x,t-\mathbf x)
    =E_0g_2^{(N)}(\mathbf x-\mathbf y)
    .
  \end{array}
  \label{hierarchy2}
\end{equation}
\end{widetext}
This equation relates $g_2$ to $g_3$ and $g_4$.
By proceeding in the same way, we can derive equations for $g_3$ and $g_4$ in terms of $g_5$ and $g_6$, and so on.
In this way, we obtain a hierarchy of equations for all the $g_n^{(N)}$.

\indent
The Full Equation is an approximation in which we truncate this hierarchy at the lowest level, by assuming that $g_3$ and $g_4$ can be expressed in terms of $g_2$, which turns\-~(\ref{hierarchy2}) into an equation for $g_2^{(N)}$ alone.
Remembering that $g_n$ can be interpreted as a correlation function, it is natural to approximate $g_3$ and $g_4$ by
\begin{equation}
  \begin{array}{>\displaystyle l}
    g^{(N)}_3(\mathbf x_2-\mathbf x_1,\mathbf x_3-\mathbf x_1)
    =\\[0.3cm]\hskip20pt=
    g^{(N)}_2(\mathbf x_2-\mathbf x_1)g^{(N)}_2(\mathbf x_3-\mathbf x_1)g^{(N)}_2(\mathbf x_3-\mathbf x_2)
  \end{array}
  \label{factor3}
\end{equation}
and
\begin{equation}
  \begin{array}{>\displaystyle l}
    g^{(N)}_4(\mathbf x_2-\mathbf x_1,\mathbf x_3-\mathbf x_1,\mathbf x_4-\mathbf x_1)
    =\\[0.3cm]\hskip20pt=
    \prod_{i<j}(g^{(N)}_2(\mathbf x_j-\mathbf x_i)+R(\mathbf x_j-\mathbf x_i))
  \end{array}
  \label{factor4}
\end{equation}
in which the correction term $R(\mathbf x_j-\mathbf x_i)=O(V^{-1})$ is relevant because $g_4^{(N)}$ appears in\-~(\ref{hierarchy2}) in a term that diverges as $V$ in the thermodynamic limit.
This correction term is computed by ensuring that $\int d\mathbf x_3d\mathbf x_4\ g_4^{(N)}=V^2g_2^{(N)}$:
\begin{equation}
  \begin{array}{>\displaystyle l}
    R(\mathbf x-\mathbf y)=
    -\frac2Vg_2^{(N)}(\mathbf x-\mathbf y)
    \cdot\\[0.3cm]\cdot
    \int d\mathbf z\ (1-g_2^{(N)}(\mathbf z-\mathbf x))(1-g_2^{(N)}(\mathbf z-\mathbf y))
    +O(V^{-2})
    .
  \end{array}
\end{equation}
Taking the thermodynamic limit $N,V\to\infty$, $\frac NV=\rho$, we find\-~(\ref{fulleq}) by defining
\begin{equation}
  g_2(\mathbf x)=:1-u(\mathbf x)
  .
\end{equation}
Furthermore, by\-~(\ref{EN}), the prediction for the ground state energy is
\begin{equation}
  \tilde e=\frac\rho2\int d\mathbf x\ (1-u(\mathbf x))v(\mathbf x)
  .
  \label{erel}
\end{equation}

\indent
The factorization assumption\-~(\ref{factor3})-(\ref{factor4}) simply states that many-body correlations of $\psi_0$ reduce to pair correlations.
If $\psi_0$ were Gaussian, this would hold exactly.
If $\psi_0$ were a Bijl-Dingle-Jastrow function \cite{Bi40,Di49,Ja55}, that is, if
\begin{equation}
  \psi_0=\prod_{i<j}e^{-\beta\varphi(\mathbf x_i-\mathbf x_j)}
  \label{jastrow}
\end{equation}
then the factorization property at long distances would be equivalent to the fact that the classical statistical mechanical model with interaction $\varphi$ satisfies the {\it clustering property}\-~\cite{Ru99}.
One can expect this to be true at low densities, where the Bijl-Dingle-Jastrow function might be a good approximation of the ground state.
At high densities, since the system approaches a mean-field regime, one might also suppose that the factorization assumption may not be so far off.
\bigskip

\indent
The Full Equation we have derived is quite difficult to study, even numerically.
As was discussed in Section\-~\ref{sec:intro}, we will introduce further approximations to simplify the equation.
The first approximation is to neglect the $\frac12u(\mathbf z)u(\mathbf y-\mathbf x)$ term in\-~(\ref{L}), which is the most difficult term, from a computational point of view. 
We expect that, at low densities, this term is expected to be of order $\rho^{3/2}$ uniformly in $\mathbf x$, whereas the leading order term in $L$ should be of order $\rho$.
This leads us to the Big Equation defined in\-~(\ref{bigeq}).
This equation is easier to solve numerically than the Full Equation, because in Fourier space, it involves only two convolution operators, whereas the Full Equation contains three, which makes it computationally heavier.
Nevertheless, this equation is still difficult to study analytically, so we make further approximations
\bigskip

\indent
Following the same idea, we can further neglect the $2u\ast(u(u\ast S))$ term in\-~(\ref{Lbigeq}).
Furthermore, we expect $u$ to decay as $|\mathbf x|^{-4}$, so if we focus on distances that are appreciably large, we can approximate $1-u$ by $1$ in the prefactor of $K$ and $L$ in\-~(\ref{bigeq}).
This leads to the Medium Equation\-~(\ref{mueq}).
\bigskip

\indent
To arrive at the Simple Equation, we take advantage of a separation of scales that occurs at low density.
On account of (\ref{EN}), the function $S(\mathbf x)$ defined in (\ref{K}) satisfies
\begin{equation}
  \int d\mathbf x\ S(\mathbf x) = \frac{2\tilde e}{\rho}
\end{equation}
which is just another way of stating (\ref{erel}).
There are two different length scales in the problem: the first is the scattering length of the potential $a_0$ and the second is the interparticle distance $\rho^{-1/3}$.
At sufficiently low densities we will have 
\begin{equation}
  a_0 \ll  \rho^{-1/3}
\end{equation}
and if  the length scale $\rho^{-1/3}$ is characteristic of the solution $u$ of (\ref{fulleq}), as we argue below, then we can expect $u(\mathbf x)$ to satisfy a bound of the form $|\nabla u(\mathbf x)| \leqslant C\rho^{1/3}$ uniformly in $\mathbf x$.
When integrating $S(\mathbf x)$ against such a slowly varying function, we may as well replace it with $2\tilde e/\rho$ times a delta function:
\begin{equation}\label{approx1}
  S(\mathbf x) \approx \frac{2\tilde e}{\rho}\delta(\mathbf x)
  .
\end{equation}
Making this approximation in\-~(\ref{K}) and\-~(\ref{Lmueq}), we arrive at the Simple Equation\-~(\ref{simpleq}).
Notice the energy per particle $\tilde e$ appears as an explicit parameter in the Simple Equation, unlike  the Full Equation.

\section{Comparison with Quantum Monte Carlo simulations}\label{sec:montecarlo}

\indent
Exact ground state properties of finite N Boson systems can be calculated arbitrarily well numerically with QMC methods.
At zero temperatures, 
it is convenient to first introduce a trial wave function, $\psi_{\mathrm{trial}}$, containing
parameters which are numerically optimized by minimizing the corresponding variation
energy evaluated by variational Monte Carlo (VMC) calculations \cite{Mc65}. 
Subsequently, the exact ground state wave function $\psi_0$ is accessed stochastically by imaginary time projection \cite{Ka70,Ce95,BM99}.
\bigskip

\indent
Here, we have performed ground state QMC calculations for $N$ bosons in a periodic box interacting
with an exponential potential, $\alpha e^{-|\mathbf{x}|}$.
Our calculations are based on a pair-product (Bijl-Dingle-Jastrow) trial wave function, 
$\psi_{\mathrm{trial}} \propto \exp(-\sum_{i<j} \varphi(|\mathbf{x}_i-\mathbf{x}_j|))$, where 
$\varphi$ is parametrized via locally piecewise-quintic Hermite interpolants in real space and 
Fourier coefficients in reciprocal space.
\bigskip

\indent
In variational Monte Carlo, $\psi_{\mathrm{trial}}^2$ is 
sampled by Metropolis Monte Carlo, and the optimal variational parameters of $\varphi$ are determined
by minimizing a linear combination of the energy and its variance. Using the optimized $\psi_{\mathrm{trial}}$ as a guiding function,
the mixed distribution $\psi_0 \psi_{\mathrm{trial}}$ is then stochastically sampled by diffusion Monte Carlo (DMC).
Linear extrapolation is used to reduce the
mixed-estimator bias occurring for observables different from the ground state energy \cite{CK86}.
In principle, the mixed-estimator bias can be controlled either by systematic improvement of
the trial wave function \cite{RMH18} or by different projection Monte Carlo methods, e.g. Reptation 
Monte Carlo \cite{BM99}. For the system under consideration, the mixed estimator bias
of the pair-product wave function was found to be sufficiently small, 
the overall precision being limited rather by
the finite system size of the QMC calculations.
\bigskip

\indent
In contrast to the computation of the Big, Medium and Simple equations, QMC calculations require an explicit numerical extrapolation from finite to infinite system size, which is frequently one of the main bottlenecks of the method.
Finite size errors in the kinetic and potential energy 
can be quantified based on two-body correlation functions \cite{HCe16}.
In addition, we have performed VMC and DMC 
calculations for various system sizes, ranging from $N=8$ to $N=512$ bosons,
to accurately extrapolate to the thermodynamic limit.
\bigskip

\indent
In the figures, errors of the QMC calculations are smaller than the size of the crosses in the plots, see Fig.\-~\ref{fig:energy}.
QMC results for hard core Bosons are taken from Ref.\-~\cite{GBC99}.

\indent

\subsection{Energy}\label{sec:energy}
\indent
Of the observables considered in this paper, the ground state energy is the most straightforward to compute: by\-~(\ref{erel}), the prediction for the energy is
\begin{equation}
  \tilde e=\int d\mathbf x\ (1-u(\mathbf x))v(\mathbf x)
  .
\end{equation}
In our notation, $e_0$ is the ground state energy per particle for the exact ground state of the Bose gas, and $\tilde e$ is the prediction for the ground state energy by the Big, Medium or Simple equation.
\bigskip

\begin{figure}
  \includegraphics[width=8cm]{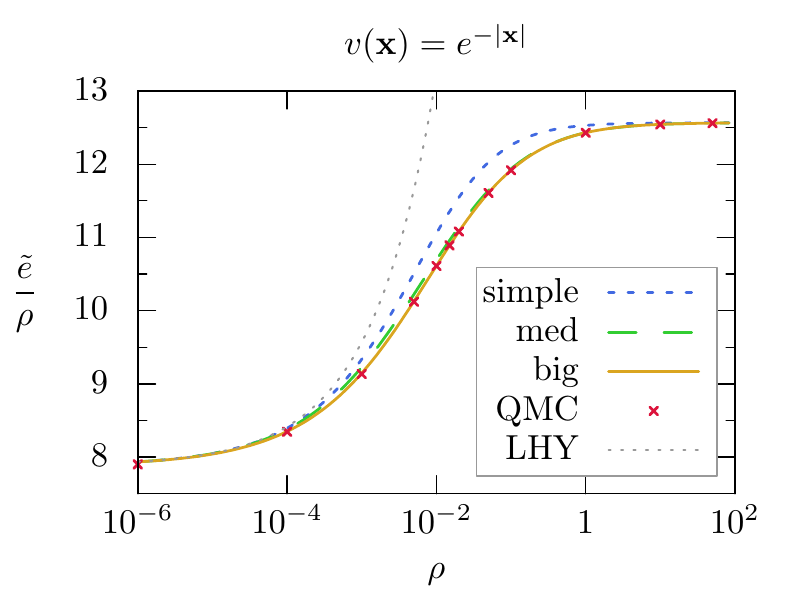}
  \hfil\includegraphics[width=8cm]{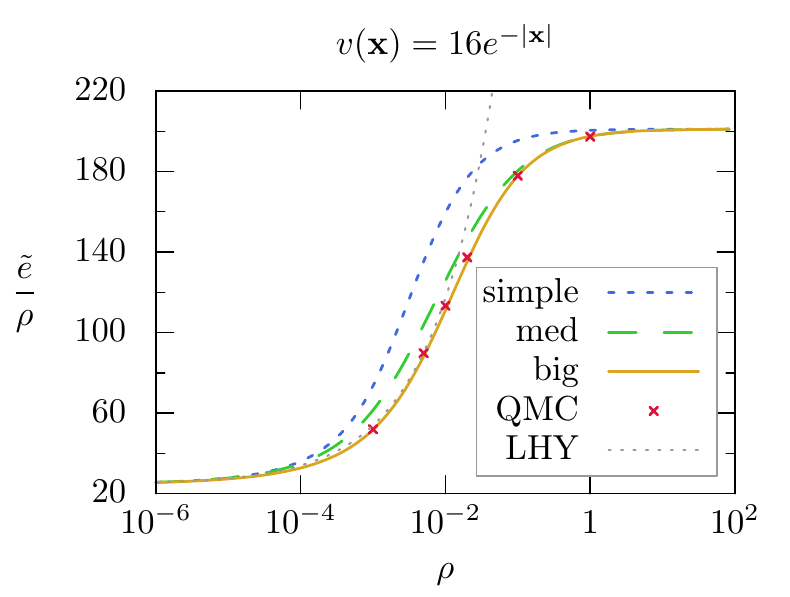}
  \caption{
    The energy as a function of density for the potential $e^{-|\mathbf x|}$ (top) and $16e^{-|\mathbf x|}$ (bottom).
    We compare the predictions of the Big, Medium and Simple Equations to a QMC simulation.
    For comparison, we also plot the Lee-Huang-Yang (LHY) energy\-~(\ref{lhy}).
    (All quantities plotted in this and the following figures are dimensionless.)
  }
  \label{fig:energy}
\end{figure}

\indent
In Figure\-~\ref{fig:energy}, we show a comparison of the prediction $\tilde e$ with a QMC simulation for the exponential potential $\alpha e^{-|\mathbf x|}$.
In\-~\cite{CJL20}, we proved that the energy prediction of the Simple Equation is asymptotically correct in both the low and high density limits.
The numerics confirm this for all three equations.
For $\alpha=1$, the Simple Equation is somewhat accurate, although the Medium and Big Equations are much closer to the QMC simulation.
For $\alpha=16$ this is even clearer, and one sees that the Medium Equation is more accurate at large densities than at small ones.
\bigskip

\begin{figure}
  \hfil\includegraphics[width=8cm]{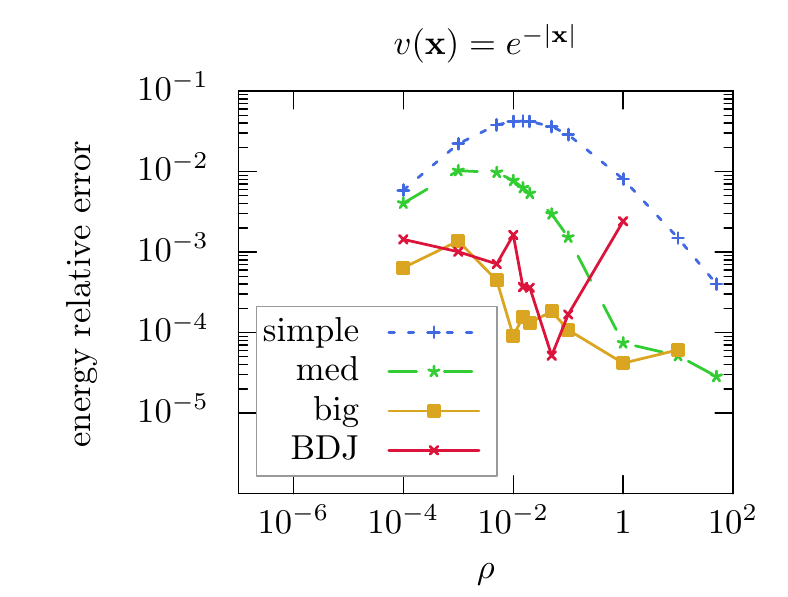}
  \hfil\includegraphics[width=8cm]{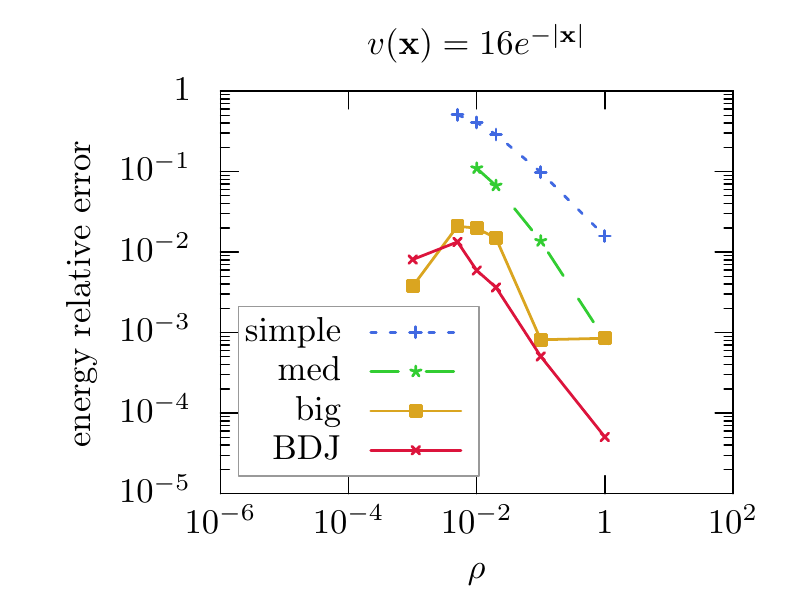}
  \caption{
    Relative error for the energy $\frac{\tilde e-e_{\mathrm{QMC}}}{e_{\mathrm{QMC}}}$ compared to the QMC simulation as a function of density for the potential $e^{-|\mathbf x|}$ (top) and $16e^{-|\mathbf x|}$ (bottom).
    The red crosses are the result for the optimal Bijl-Dingle-Jastrow (BDJ) function.
  }
  \label{fig:cmp_energy}
\end{figure}

\indent
A more quantitative comparison can be found in Figure\-~\ref{fig:cmp_energy}, where we plot the relative error, that is, $(\tilde e-e_{\mathrm{QMC}})/e_{\mathrm{QMC}}$, where $e_{\mathrm{QMC}}$ is the Quantum Monte-Carlo prediction for the energy.
We find that, for $\alpha=1$, the relative error is, at most, 5\% for the Simple Equation, 1\% for the Medium Equation, and $0.1\%$ for the Big Equation.
For $\alpha=16$, all equations are less accurate, with a relative error of 60\% for the Simple Equation, 10\% for the Medium Equation, and 2\% for the Big Equation.

\indent
In addition, in Figure\-~\ref{fig:cmp_energy}, we compare with the error made by the optimal Bijl-Dingle-Jastrow function.
A Bijl-Dingle-Jastrow function is an Ansatz for the ground state wave function of the form\-~(\ref{jastrow}).
Finding the optimal function $\varphi$ which minimizes the energy is a computationally intensive operation, which is used as a first approximation when running the diffusion QMC simulation used in Figure\-~\ref{fig:energy}.
We find that the optimal Bijl-Dingle-Jastrow function gives a prediction for the ground state energy which is about as accurate as the Big Equation.
Of note is the fact that solving the Big Equation numerically is computationally much less difficult than computing the optimal Bijl-Dingle-Jastrow function.
In addition, in Figure\-~\ref{fig:cmp_energy}, we see that the Full Equation and the Big Equation produce very similar results.

\subsection{Condensate fraction}
\indent
The approximations leading to the Big, Simple and Medium Equations reduce the number of degrees of freedom from $3N$ in the many body Bose gas to just $3$.
In doing so, we lose some information, and, in particular, we do not obtain a prediction for the many-body wavefunction $\psi_0$.
Therefore, computing observables other than the ground state energy is not entirely straightforward.
To compute the condensate fraction, we first express it in terms of the energy of an auxiliary system, from which we derive an approximation following the prescriptions in section\-~\ref{sec:approx}.
Specifically, the {\it non}-condensed fraction of the many-body ground state $\psi_0$
\begin{equation}
  \eta_0:=1-\frac1N\sum_{i=1}^N\left<\psi_0\right|P_i\left|\psi_0\right>
\end{equation}
is expressed in terms of the projector $P_i\psi_0:=\int\frac{d\mathbf x_i}V\psi_0$ onto the condensate wavefunction (which is the constant function):
which we re-express in terms of the modified Hamiltonian
\begin{equation}
  H_\mu=-\frac12\sum_{i=1}^N\Delta_i+\sum_{1\leqslant i<j\leqslant N}v(\mathbf x_i-\mathbf x_j)-\mu\frac1N\sum_{i=1}^NP_i
\end{equation}
whose ground state energy per particle is denoted by $e_{0,\mu}$:
\begin{equation}
  \eta_0=1+\left.\partial_\mu e_{0,\mu}\right|_{\mu=0}
  .
\end{equation}
Following the approximation scheme in section\-~\ref{sec:approx}, we compute an approximation $\tilde e_\mu$ for $e_{0,\mu}$ (following the convention used before, $e_{0,\mu}$ is the energy for the exact many-body ground state and $\tilde e_\mu$ is the prediction of the Big, Medium and Simple equations):
\begin{equation}
  (-\Delta+2\mu) u_\mu(\mathbf x)
  =
  (1-u_\mu(\mathbf x))\left(v(\mathbf x)-2\rho K(\mathbf x)+\rho^2 L(\mathbf x)\right)
\end{equation}
\begin{equation}
  \tilde e_\mu=\int d\mathbf x\ (1-u_\mu(\mathbf x))v(\mathbf x)
\end{equation}
(compare this to\-~(\ref{fulleq})).
This leads to an approximation $\tilde\eta$ for the non-condensed fraction $\eta_0$:
\begin{equation}
  \tilde\eta:=1+\partial_\mu\tilde e_\mu|_{\mu=0}
  .
\end{equation}
Proceeding as in section\-~\ref{sec:approx}, we obtain predictions for the Big, Simple and Medium Equations.
\bigskip

\indent
In the case of the Simple Equation, we can relate $\tilde\eta$ and the solution $u$ of the equation\-~(\ref{simpleq}) directly:
\begin{equation}
  \tilde\eta=\frac{\int d\mathbf x\ v(\mathbf x)\mathfrak K_{\tilde e}u(\mathbf x)}{1-\rho\int d\mathbf x\ v(\mathbf x)\mathfrak K_{\tilde e}(2u(\mathbf x)-\rho u\ast u(\mathbf x))}
\end{equation}
where $\mathfrak K_{\tilde e}$ is the operator
\begin{equation}
  \mathfrak K_{\tilde e}:=(-\Delta+4\tilde e(1-\rho u\ast)+v)^{-1}
  .
  \label{Ke}
\end{equation}
In\-~\cite{CJL20b}, we study this operator in detail, and derived the low density limit of $\tilde\eta$:
\bigskip
\begin{equation}
  \tilde\eta\mathop\sim_{\rho\to0}\frac{8\sqrt{\rho a_0^3}}{3\sqrt\pi}
  \label{eta_asym}
\end{equation}
which agrees with the prediction of Bogolubov theory\-~(\ref{eta0})\-~\cite[(41)]{LHY57}.
\bigskip

\begin{figure}
  \hfil\includegraphics[width=8cm]{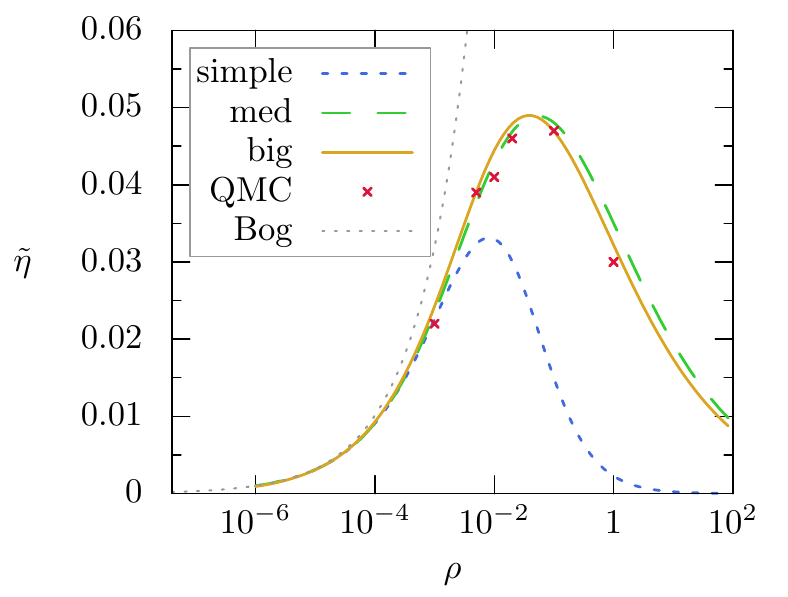}
  \caption{
    The non-condensed fraction as a function of the density for the potential $\frac12e^{-|\mathbf x|}$.
    We compare the predictions of the Big, Medium and Simple Equations to a QMC simulation.
    The prediction of Bogolubov theory\-~(\ref{eta0}) is also plotted for comparison (Bog).
  }
  \label{fig:condensate0.5}
\end{figure}

\indent
For the Big and Medium Equation, we carried out numerical computations, the results of which are reported in Figure\-~\ref{fig:condensate0.5}.
Whereas all three approximate equations agree with one another very well at low densities, the Simple Equation becomes less accurate at intermediate densities.
However, the Big and Medium Equations make rather accurate predictions (though not as accurate as for the energy), compared to the QMC simulation.
We find, as expected, that all particles are condensed both at zero density and at infinite density, where the Bose gas becomes a mean-field system.
The location of the maximum of the non-condensed fraction (or the minimum of the condensed fraction) is accurately predicted by the Big and Medium Equations.

\subsection{Two-point correlation function}\label{sec:2pt}
\indent
The two-point correlation function in the ground state is defined as
\begin{equation}
  C_2(\mathbf y-\mathbf y'):=\sum_{i,j=1}^N\left<\psi_0\right|\delta(\mathbf y-\mathbf x_i)\delta(\mathbf y'-\mathbf x_j)\left|\psi_0\right>
  .
\end{equation}
We first note that this can be rewritten in a way that makes the translation invariance of $C_2$ more apparent, by denoting $\mathbf x:=\mathbf y-\mathbf y'$ and taking an average over $\mathbf y'$:
\begin{equation}
  C_2(\mathbf x):=\frac2V\sum_{1\leqslant i<j\leqslant N}\left<\psi_0\right|\delta(\mathbf x-(\mathbf x_i-\mathbf x_j))\left|\psi_0\right>
\end{equation}
which we can rewrite as a functional derivative of the ground state energy per-particle $e_0$:
\begin{equation}
  C_2(\mathbf x)=2\rho^2\frac{\delta e_0}{\delta v(\mathbf x)}
  .
\end{equation}
The prediction $\tilde C_2$ of the Big and Medium Equations for the two-point correlation function are therefore defined by differentiating $\tilde e$ in\-~(\ref{erel}) with respect to $v$:
\begin{equation}
  \tilde C_2(\mathbf x):=2\rho^2\frac{\delta\tilde e}{\delta v(\mathbf x)}
  .
  \label{C2}
\end{equation}

\indent
In the case of the simple equation, we will proceed differently.
If we were to define $\tilde C_2$ as in\-~(\ref{C2}), we would find that $\tilde C_2$ would not converge to $\rho^2$ as $|\mathbf x|\to\infty$, which is obviously unphysical.
This comes from the fact that first approximating $S$ as in\-~(\ref{approx1}) and then differentiating with respect to $v$ is less accurate than first differentiating with respect to $v$ and then approximating $S$.
Defining $\tilde C_2$ following the latter prescription, we find that, for the Simple Equation,
\begin{equation}
  \begin{array}{>\displaystyle l}
    \tilde C_2(\mathbf x)=
    \rho^2\tilde g_2(\mathbf x)+
    \\[0.3cm]
    +\rho^2\frac{\mathfrak K_{\tilde e}v(\mathbf x)\tilde g_2(\mathbf x)-2\rho u\ast \mathfrak K_{\tilde e}v(x)+\rho^2u\ast u\ast \mathfrak K_{\tilde e}v(x)}{1-\rho\int d\mathbf x\ v(\mathbf x)\mathfrak K_{\tilde e}(2u(\mathbf x)-\rho u\ast u(\mathbf x))}
  \end{array}
  \label{correlation_simpleq}
\end{equation}
where $\mathfrak K_{\tilde e}$ is the operator defined in\-~(\ref{Ke}).
Defined in this way, $\tilde C_2\to\rho^2$ as $|\mathbf x|\to\infty$.
\bigskip

\indent
$C_2$ is the physical correlation function, using the probability distribution $|\psi_0|^2$, but, as we saw in section\-~\ref{sec:approx}, $\psi_0$ can also be thought of a probability distribution, whose two-point correlation function is $g_2$, defined in\-~(\ref{g}).
The Big, Medium and Simple Equations make a natural prediction for the function $g_2$: namely $1-u(\mathbf x)$.
\bigskip

\begin{figure}
  \hfil\includegraphics[width=8cm]{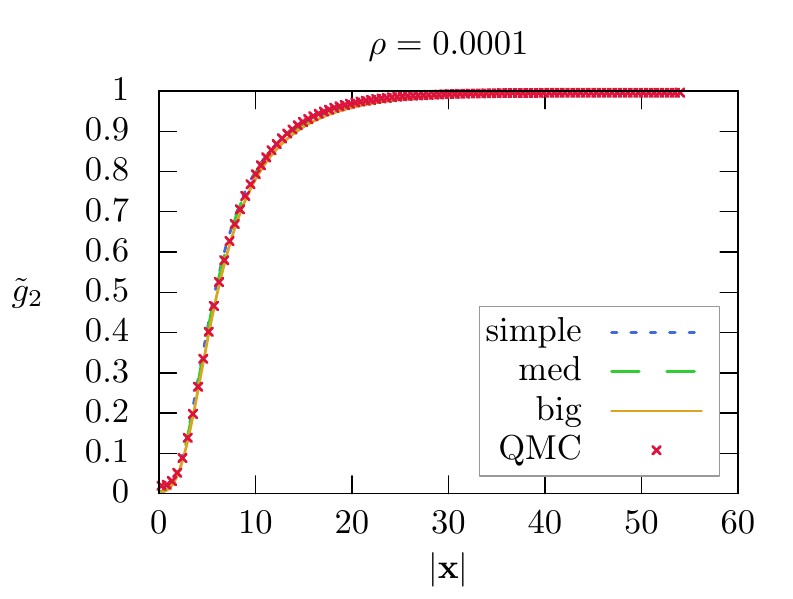}
  \hfil\includegraphics[width=8cm]{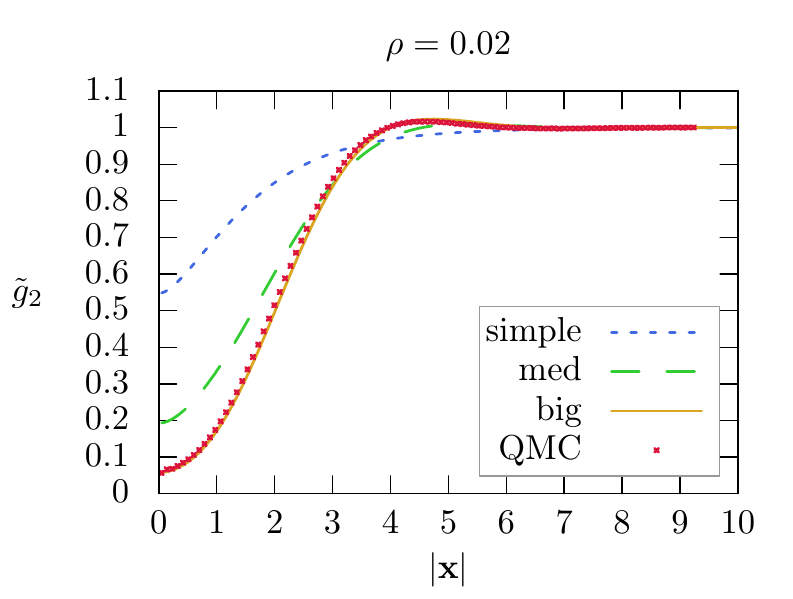}
  \caption{
    $\tilde g_2(\mathbf x)$ for the potential $16e^{-|\mathbf x|}$ at $\rho=0.0001$ (top) and $\rho=0.02$ (bottom).
    We compare the predictions of the Big, Medium and Simple Equations to a QMC simulation.
    }
  \label{fig:ux}
\end{figure}

\indent
In Figure\-~\ref{fig:ux}, we compare the prediction $\tilde g_2$ produced by the Big, Medium and Simple Equations to the QMC simulation.
We find that for low enough densities, the three predictions are consistent with one another, and accurately reproduce the result of the QMC simulation.
However, as the density is increased, there is a transition to a situation in which the predictions from the Big, Medium and Simple Equations start to differ significantly from one another.
In particular, in the case of the Simple Equation, $\tilde g_2\leqslant 1$, whereas for the Big and the Medium Equations, $\tilde g_2$ has a maximum that is $>1$.
The prediction of the Big Equation remains quite accurate, when compared to the QMC simulation, which also exhibits a bump in $g_2$.
The presence of this local maximum in $g_2$ shows that, in the probability distribution $\psi_0$, there is a larger probability of finding pairs of particles that are separated by a certain fixed distance.
This indicates the appearance of a new physical length scale at intermediate densities, and indicates that the system exhibits a non-trivial physical behavior in this regime.
Note that this behavior was observed for the stronger potential $16e^{-|\mathbf x|}$, but seems to be absent for $e^{-|\mathbf x|}$.
Note, also, that, as will be discussed next, this maximum is also present in the two-point correlation $C_2$, and is, therefore, the manifestation of a physical phenomenon.
\bigskip

\begin{figure}
  \hfil\includegraphics[width=8cm]{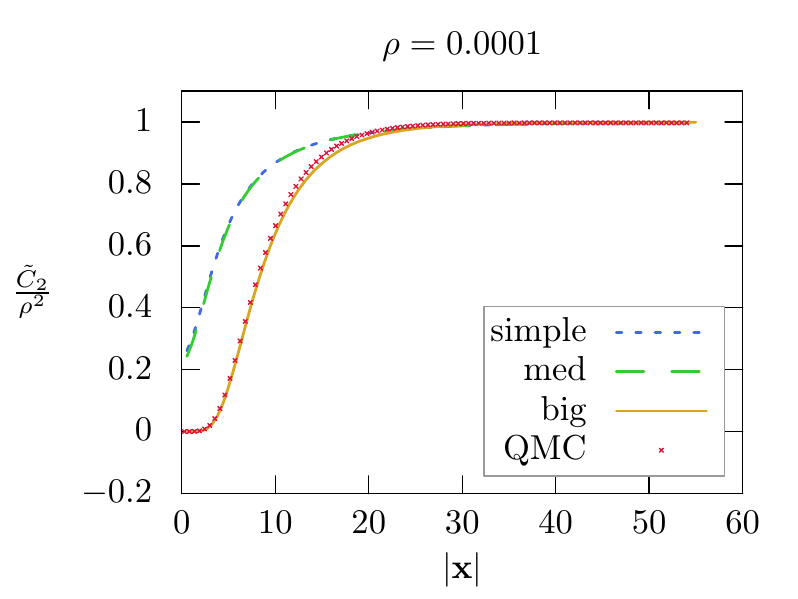}
  \hfil\includegraphics[width=8cm]{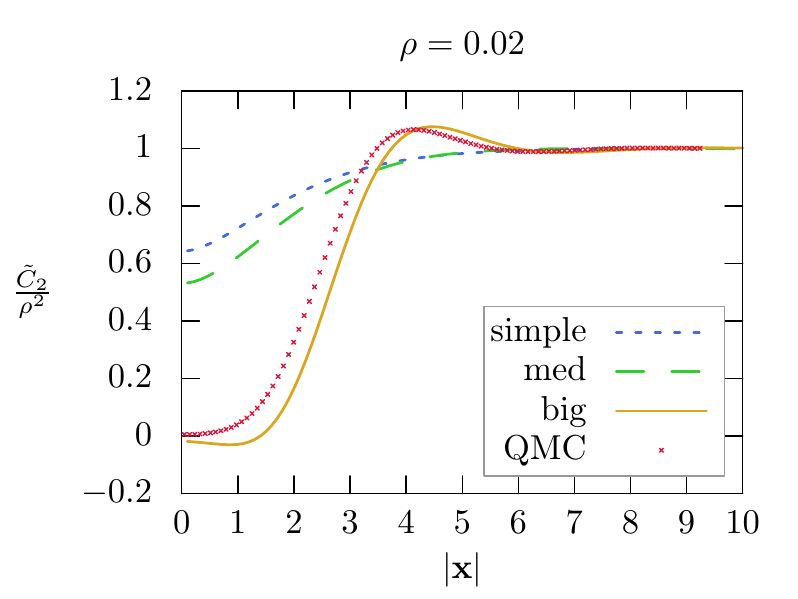}
  \caption{
    $\frac{\tilde C_2}{\rho^2}$ for the potential $e^{-|\mathbf x|}$ at $\rho=0.0001$ (top) and $\rho=0.02$ (bottom).
    We compare the predictions of the Big, Medium and Simple Equations to a QMC simulation.
  }
  \label{fig:correlation}
\end{figure}

\indent
In Figure\-~\ref{fig:correlation}, we compare the prediction $\tilde C_2$ to the QMC simulation.
At low densities, the prediction of the Big Equation agrees rather well with the QMC simulation.
The Simple and Medium Equations are not as accurate.
At larger densities, the Simple and Medium Equations are quite far from the QMC computation, and the Big Equation is not as accurate as in the case of $\tilde g_2$, but it does reproduce some of the qualitative behavior of the QMC computation.
In particular, there is a local maximum in the two-point correlation function, which occurs at a length scale that is close to that observed for $\tilde g_2$.
This suggests the emergence of a non-trivial phase, which resembles a liquid.
At small $\mathbf x$, $\tilde C_2$ is negative, which is clearly not physical, and those values should be discarded.

\subsection{Momentum distribution}
\indent
Next, we study the momentum distribution $\mathcal M_0(\mathbf k)$.
Computations carried out for the contact Hamiltonian\-~\cite{CAL09,NE17} suggest that $\mathcal M_0$ should satisfy the asymptotic relation\-~(\ref{tan})
\begin{equation}
  \mathcal M_0(\mathbf k)\sim\frac{c_2}{|\mathbf k|^4}
  ,\quad
  c_2=8\pi a_0^2\frac{\partial e_0}{\partial a}
\end{equation}
and we will now discuss whether this holds for the Big, Simple and Medium Equations.
To compute a prediction for the momentum distribution, we proceed in the same way as for the condensate fraction above.
First of all, the momentum distribution is defined as
\begin{equation}
  \mathcal M_0(\mathbf k):=\frac1N\sum_{i=1}^N\left<\psi_0\right|F_i(\mathbf k)\left|\psi_0\right>
\end{equation}
where $F_i$ is the projection onto the state $e^{i\mathbf k\mathbf x_i}$.
Thus, defining a modified Hamiltonian,
\begin{equation}
  H_\lambda=-\frac12\sum_{i=1}^N\Delta_i+\sum_{1\leqslant i<j\leqslant N}v(\mathbf x_i-\mathbf x_j)+\lambda\frac1N\sum_{i=1}^NF_i
\end{equation}
whose ground state energy per particle is denoted by $e_{0,\lambda}(\mathbf k)$:
\begin{equation}
  \mathcal M_0(\mathbf k)=\left.\partial_\lambda e_{0,\lambda}(\mathbf k)\right|_{\lambda=0}
  .
\end{equation}
Proceeding as in section\-~\ref{sec:approx}, this implies the following definition for the modified Full Equation (compare to\-~(\ref{fulleq})): for $\mathbf k\neq0$,
\begin{equation}
  \begin{array}{>\displaystyle l}
    (-\Delta+v(\mathbf x))u_\lambda(\mathbf x)
    =v(\mathbf x)-
    \\[0.3cm]\hskip20pt
    -\rho(1-u_\lambda(\mathbf x))(2K(\mathbf x)-\rho L(\mathbf x))-2\lambda\hat u(\mathbf k)\cos(\mathbf k\mathbf x)
  \end{array}
\end{equation}
where $\hat u(\mathbf k)$ is the Fourier transform of $u|_{\lambda=0}$, and
\begin{equation}
  \tilde e_\lambda(\mathbf k)=\int d\mathbf x\ (1-u_\lambda(\mathbf x))v(\mathbf x)
  .
\end{equation}
The prediction $\tilde{\mathcal M}$ for the momentum distribution $\mathcal M_0$ is then
\begin{equation}
  \tilde{\mathcal M}(\mathbf k):=\partial_\lambda\tilde e_\lambda(\mathbf k)|_{\lambda=0}
  .
\end{equation}
\bigskip

\indent
We showed in\-~\cite{CJL20b} that, in the case of the Simple Equation, (\ref{tan}) holds in the limit in which $|\mathbf k|,\rho\to0$ while $\frac{|\mathbf k|}{2\sqrt{\tilde e}}\to\infty$.
This suggests that the Tan relation\-~(\ref{tan}) only holds in the range
\begin{equation}
  \sqrt\rho\ll|\mathbf k|\ll1
\end{equation}
and, in particular, that if $\sqrt\rho\gtrsim1$, then the Tan relation does not hold at all, which means that the physics of the Bose gas at intermediate densities is of a different nature from that studied in the context of the unitary Bose gas.
\bigskip

\begin{figure}
  \hfil\includegraphics[width=8cm]{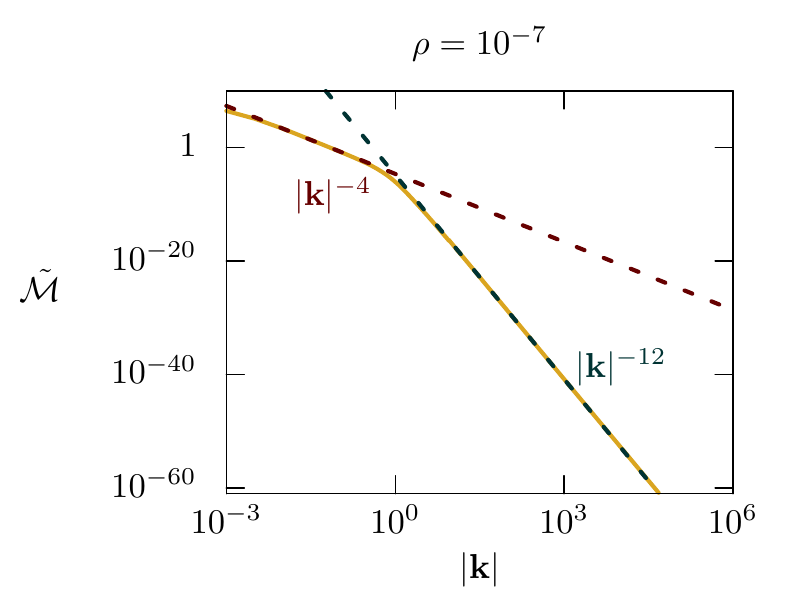}
  \hfil\includegraphics[width=8cm]{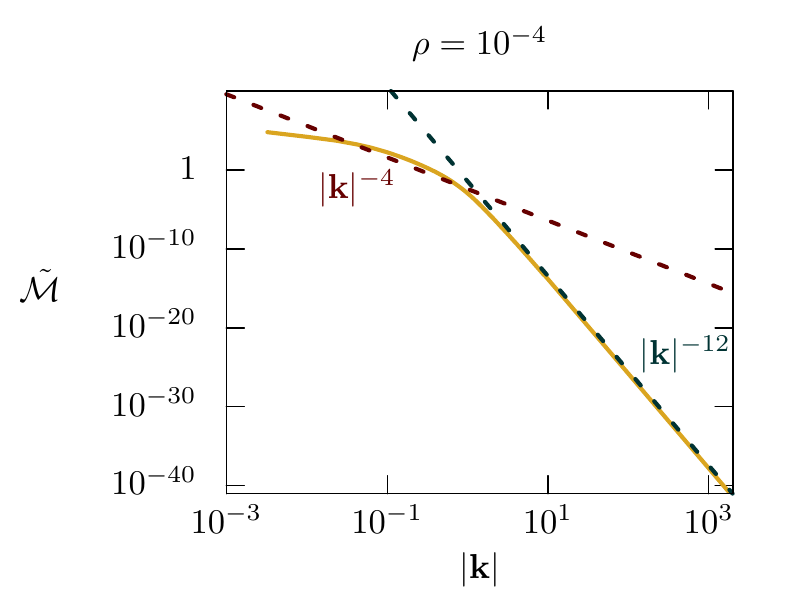}
  \caption{
    The prediction of the Big Equation for the momentum distribution as a function of $|\mathbf k|$ for the potential $e^{-|\mathbf x|}$, $\rho=10^{-7}$ (top) and $\rho=10^{-4}$ (bottom).
    The dark red dotted line has a slope of $-4$ and corresponds to a $|\mathbf k|^{-4}$ behavior, whereas the dark green dotted line has a slope $-12$, and corresponds to $|\mathbf k|^{-12}$.
  }
  \label{fig:tan}
\end{figure}

\indent
In Figure\-~\ref{fig:tan}, we show a numerical computation of $\tilde{\mathcal M}(\mathbf k)$ for the Big Equation, at a very low density, and a larger one.
As was predicted for the Simple Equation, we find that the Tan universal relation\-~(\ref{tan}) holds at low density, provided $|\mathbf k|$ is small enough.
At larger values of $|\mathbf k|$, the decay of $\hat v(\mathbf k)$ kicks in, and the momentum distribution decays much faster.
As the density is increased, the domain in which $\tilde{\mathcal M}(\mathbf k)\sim|\mathbf k|^{-4}$ shrinks to nothing, and the Tan universal relation completely disappears.
\bigskip

\indent
Here, we have not attempted a direct comparison of the momentum distribution with QMC calculations.
From the previous comparisons of the energy, pair correlations, and condensate fraction, we expect that, at the two densities considered in Figure\-~\ref{fig:tan}, the deviation of the prediction of the Big Equation from the exact ground state are expected to be smaller than the stochastic error limiting the precision of QMC calculations of the momentum distribution.
This is particularly true in the region in which $|\mathbf k|^{-4}$ transitions to $|\mathbf k|^{-12}$.

\subsection{Non-universal behavior at intermediate densities}
\indent
The low density asymptotics of the energy, given by the Lee-Huang-Yang formula\-~(\ref{lhy}), only depend on the potential through the scattering length.
At high density\-~(\ref{ehigh}), they only depend on the potential through $\int d\mathbf x\ v(\mathbf x)$.
In this sense, the low and high density behavior of the Bose gas is {\it universal}.
In this section, we show that, at intermediate densities, the energy does not only depend on the scattering length and the integral of the potential, thus suggesting that the behavior of the Bose gas at intermediate densities is {\it not universal}.

\indent
To that end, we have compared the predictions of the Big Equation for the energy for two potentials that have the same scattering length, and the same integral.
The first potential, $v_{32}^{(0)}$, is defined in the next section, see\-~(\ref{vn}), and the second is an exponential potential
\begin{equation}
  \Phi_{\alpha,\beta}(\mathbf x):=\alpha e^{-\beta|\mathbf x|}
\end{equation}
where $\alpha$ and $\beta$ are chosen in such a way that the scattering length and integral of $\Phi$ are equal to those of $v_{32}^{(0)}$.
The scattering length of $v_{32}^{(0)}$ was computed numerically and found to be $\approx0.5878$, and its integral is $\frac{64\pi^2}9$.
The scattering length of $\Phi_{\alpha,\beta}$ is
\begin{equation}
  \frac1\beta\left(\log\frac\alpha{\beta^2}+2\gamma+2\frac{K_0(2\sqrt{\frac\alpha{\beta^2}})}{I_0(2\sqrt{\frac\alpha{\beta^2}})}\right)
\end{equation}
where $\gamma$ is the Euler constant and $K_0$ and $I_0$ are modified Bessel functions.
The integral of $\Phi_{\alpha,\beta}$ is $\frac{8\pi\alpha}{\beta^3}$.
We thus find that, in order to make the scattering length and integral of $v_{32}^{(0)}$ and $\Phi_{\alpha,\beta}$ coincide, we must choose $\alpha\approx907.2$ and $\beta\approx6.874$.

\begin{figure}
  \hfil\includegraphics[width=8cm]{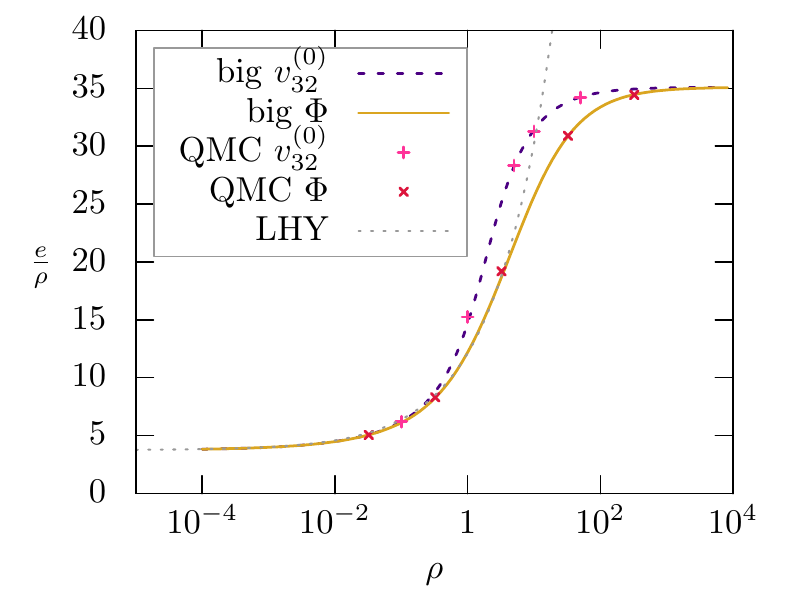}
  \caption{
    The prediction of the energy by the Big Equation for the potentials $v_{32}^{(0)}$ and $\Phi(\mathbf x)\equiv\alpha e^{-\beta|\mathbf x|}$ with $\alpha\approx 907.2$ and 
$\beta \approx 6.873$.
    The potentials are chosen to have the same scattering length, $a_0\approx 0.5878$, as well
as the same value for their integrals, so they coincide at low and at high densities.
    They differ signinficantly at intermediate densities.
    We compare each curve to a few QMC points, which fit well.
    We also plot the Lee-Huang-Yang (LHY) energy\-~(\ref{lhy}).
  }
  \label{fig:compare_pots}
\end{figure}

\indent
The prediction of the energy for these two potentials is plotted in Figure\-~\ref{fig:compare_pots}.
We find that, as expected, the energies coincide at low and high density, but they differ significantly in the intermediate density regime.
We have confirmed this fact by QMC computations, and found good agreement of the QMC data with our prediction for both potentials.

\section{Hard-core potential}\label{sec:hardcore}

\indent
The numerical computations discussed above as well as the proofs in\-~\cite{CJL20,CJL20b} heavily use the assumption that the potential $v$ is integrable, which a priori excludes the case of a hard-core potential, which is infinite inside a radius $1$.
We have investigated two directions to get around this restriction.

\indent
The first, and most straightforward, is to consider the hard-core potential as a limit of soft core potentials.
Obviously, this approach will not be accurate at densities approaching close-packing, but as we will see, is rather accurate at smaller densities.
As was mentioned in section\-~\ref{sec:intro}, it is preferable to only use potentials of positive type (that is, non-negative potentials with a non-negative Fourier transform).
With this in mind, we consider the sequence of potentials
\begin{equation}
  v^{(0)}_n(|\mathbf x|)
  :=\Theta(1-|\mathbf x|)
  \alpha_n\frac{2\pi}{3}(|\mathbf x|-1)^2(|\mathbf x|+2)
\end{equation}
where $\Theta(x)$ is the Heaviside function, which is equal to $1$ for $x>0$ and $0$ otherwise, and $\alpha_n\to\infty$.
This potential can also be written as
\begin{equation}
  v^{(0)}_n(|\mathbf x|)=
  \alpha_n\int d\mathbf y\ \Theta({\textstyle\frac12-|\mathbf y|})\Theta({\textstyle\frac12-|\mathbf x-\mathbf y|})
\end{equation}
which shows that it is of positive type because it is the convolution of the function $\Theta(\frac12-|\mathbf x|)$ with itself.
In addition, we fix the scattering length of the potential to 1, by rescaling space: denoting the scattering length of $v^{(0)}_n$ by $a_n$, we take the potential to be
\begin{equation}
  v_n(\mathbf x):=v^{(0)}_n\left({\textstyle\frac{|\mathbf x|}{a_n}}\right)
  .
  \label{vn}
\end{equation}

\indent
The second method is to solve the Big, Medium and Simple Equations for $|\mathbf x|>1$, with the boundary condition $u(\mathbf x)=1$ at $|\mathbf x|=1$.
From a computational standpoint, the Big and Medium Equations were too difficult to solve quickly on our hardware.
In the case of the Simple Equation, the computation is much longer than in the case of a soft-core potential, but it is not excessively long.
The reason for which solving the equation for $|\mathbf x|>1$ is computationally much more difficult than the soft core case, is that in the latter case, we carry out the computation in Fourier space, in which the Big, Simple and Medium Equations have fewer integrals.
For the hard-core potential, the Fourier transform of $u$ does not decay fast enough for the numerics to be precise, so we work in real space instead, which is computationally more difficult.
\bigskip

\begin{figure}
  \hfil\includegraphics[width=8cm]{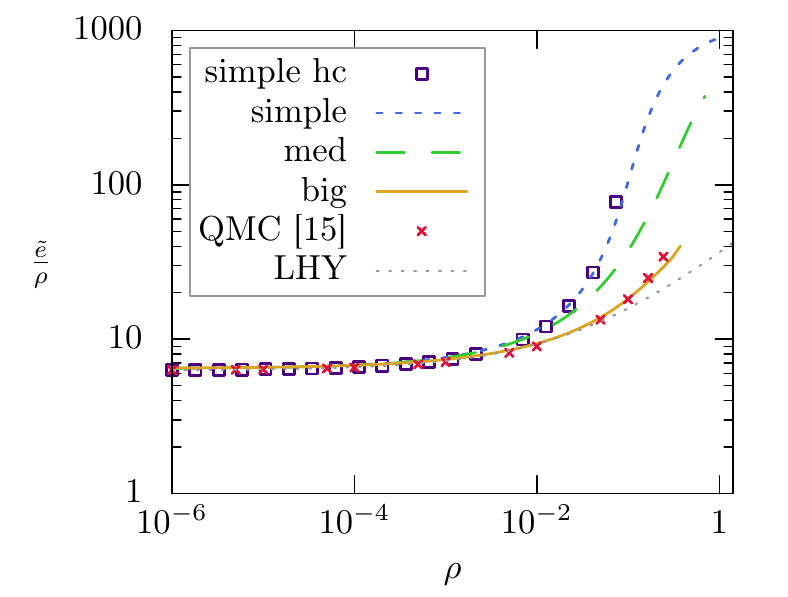}
  \hfil\includegraphics[width=8cm]{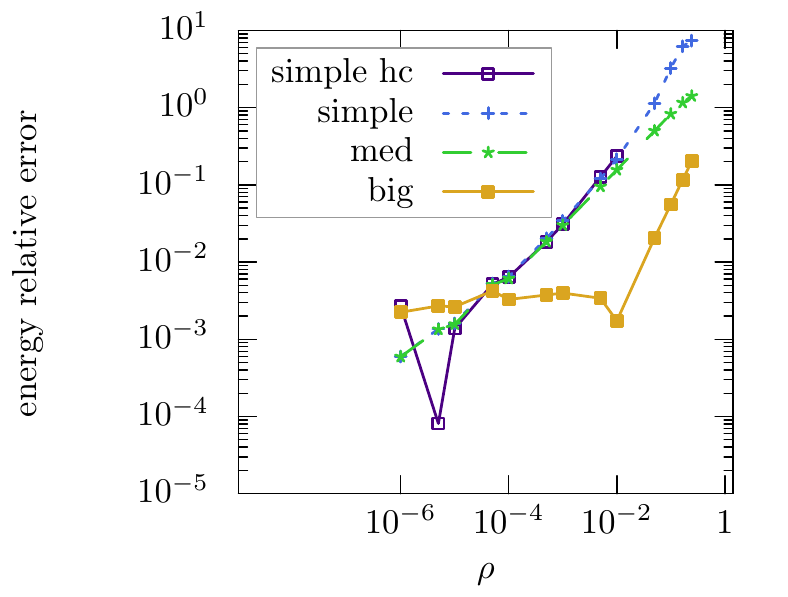}
  \hfil\includegraphics[width=8cm]{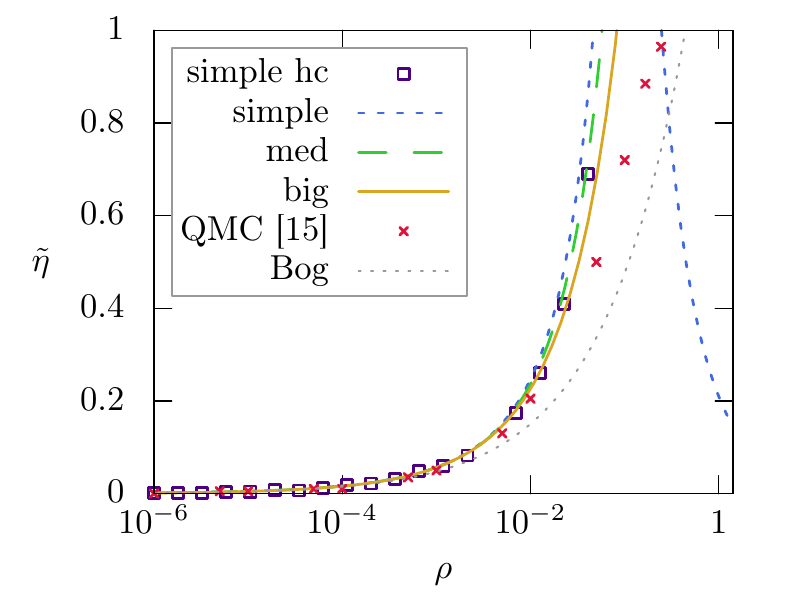}
  \caption{
    The energy (top), relative error in the energy $\frac{\tilde e-e_{\mathrm{QMC}}}{e_{\mathrm{QMC}}}$ (middle), and non-condensed fraction (bottom) as a function of the density for the hard core potential.
    The circles were computed by solving the hard core Simple Equation for $|\mathbf x|>1$ (simple hc).
    The lines were computed by approximating the hard core potential by the potential $v_{512}(\mathbf x)$, see\-~(\ref{vn}).
    We compare the predictions of the Big, Medium and Simple Equations to QMC results reported\-~\cite{GBC99}.
    The prediction of Bogolubov theory\-~(\ref{eta0}) is also plotted for comparison (Bog).
    The right edge of the plots correspond to the close-packing density $\rho_{\mathrm{cp}}=\sqrt2$\-~\cite{Ha05}.
  }
  \label{fig:hardcore}
\end{figure}

\indent
In Figure\-~\ref{fig:hardcore}, we compare the predictions for the energy and condensate fraction made using the Big, Medium and Simple Equations to the QMC computation carried out in\-~\cite{GBC99}.
The plots are shown for densities up to the close packing density, which is the maximal allowed density for the hard core potential.
All three Equations are quite accurate at low density, but the error becomes larger as the density in ramped up.
Nevertheless, for the energy, the Big Equation stays quite close to the QMC simulation.
As the density approaches close packing, the potential $v_n$ becomes inadequate.
The effects of this are most visible for the Simple Equation.
For smaller densities, for the Simple Equation, we see that the predictions made using $v_n$ are rather close to those made by restricting the equation to $|\mathbf x|>1$.
\bigskip

\section{Limits of validity of the Simple Equations}\label{sec:limits}

\indent
As we have seen above, the Big, Medium and Simple Equations are, in some cases very accurate (especially the Big Equation).
In this section, we discuss the situations in which these equations make predictions that are far from the QMC simulations, or even unphysical.
\bigskip

\indent
First of all, the Big, Medium and Simple Equations are only accurate at high densities if the potential is of positive type, that is, if its Fourier transform is non-negative.
Indeed, as we proved for the Simple Equation in\-~\cite{CJL20} and as the numerics show for the Big and Medium Equations, as $\rho\to\infty$, $\tilde e\sim\frac\rho2\int d\mathbf x\ v(\mathbf x)$.
For the Bose gas, this was proved to hold if $v$ is of positive type\-~\cite{Li63}.
It is quite easy to find a counter-example if $v$ is not of positive type.
For instance, if $v(\mathbf x)=0$ for all $|\mathbf x|<1$, then, consider a wavefunction $\psi$ that is smooth and supported on $|\mathbf x_1|,\cdots,|\mathbf x_N|<\frac 12$.
Since all particles are at a distance that is $<1$, the potential energy of such a wavefunction is 0, and its kinetic energy is $O(N)$.
Thus, the energy per particle is of order 1, which, for large $\rho$, is $\ll\frac\rho2\int d\mathbf x\ v(\mathbf x)$.
(Note that a non-trivial, non-negative potential with $v(\mathbf x)$ cannot be of positive type if $v(0)=0$, since the maximum of a positive type function is attained at $0$.)
\bigskip

\begin{figure}
  \hfil\includegraphics[width=8cm]{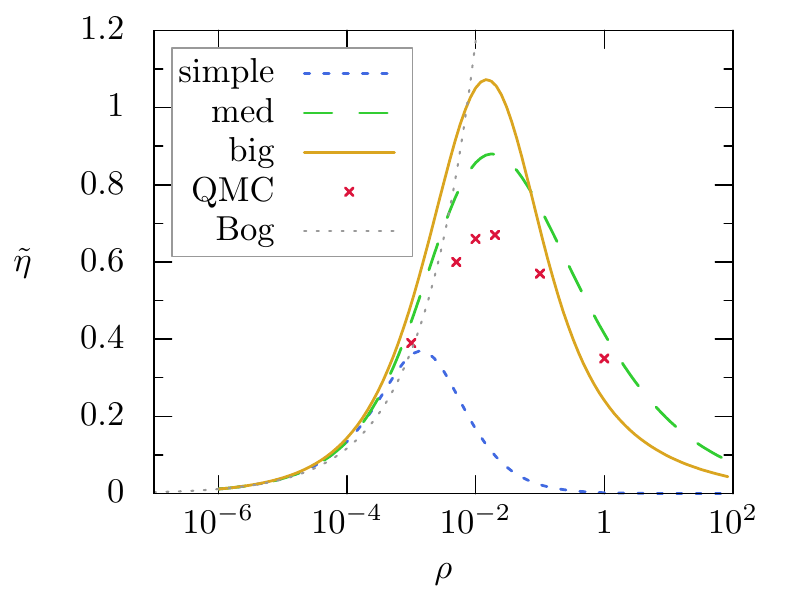}
  \caption{
    The non-condensed fraction as a function of the density for the potential $16e^{-|\mathbf x|}$.
    We compare the predictions of the Big, Medium and Simple Equations to a QMC simulation.
    The prediction of Bogolubov theory\-~(\ref{eta0}) is also plotted for comparison (Bog).
  }
  \label{fig:condensate16}
\end{figure}

\indent
In addition, we observed that the predictions made by the Big, Medium and Simple Equations get less accurate if the potential is made stronger.
Comparing the relative error in Figure\-~\ref{fig:cmp_energy} between the potential $e^{-|\mathbf x|}$ and $16e^{-|\mathbf x|}$ shows that the error is roughly 10 times worse.
For the condensate fraction, the situation deteriorates further, as can be seen in Figure\-~\ref{fig:condensate16}, in which we see that, even though the Big Equation still reproduces the qualitative features of the condensate fraction curve, it yields an unphysical result, with a negative condensate fraction.
This is further confirmed by the computations for the hard core potential, in which we see from Figure\-~\ref{fig:hardcore} that the condensate fraction becomes rather inaccurate at large densities.

\section{Conclusions}\label{sec:conclusions}

\indent
In this paper we show the good agreement in the predictions of the ground state energy, condensate fraction and correlation function of the repulsive Bose gas given by the {\it simplified approach} developed in 1963\-~\cite{Li63} with the values obtained by Quantum Monte-Carlo calculations, for the potentials $e^{-|\mathbf x|}$ and $16e^{-|\mathbf x|}$.
The simplified approach was thought to be accurate  only at low densities, in complete agreement with other analyses of the time.
Here, we show that it is accurate at {\it all} densities.
This establishes a new approach to many body bosonic physics.
Combining this analysis with the exact results in\-~\cite{CJL20,CJL20b} leads us to conjecture that the simplified approach is accurate for any repulsive potential of positive type with a scattering length and an integral that is not too large.
\bigskip

\indent
We have discussed three different approximations, the Big, Medium and Simple Equations.
The Big Equation is the most accurate, but also the most difficult to solve.
The Medium Equation is obtained by neglecting terms of higher order in $u$, which makes it much more easy to compute with, while remaining rather close to the Big Equation.
The Simple Equation is then obtained by approximating $g_2(x)v(x)$ by a Dirac-delta function.
This drastically simplifies the equation, but is also less accurate at intermediate densities (while the low and high densities are still asymptotically exact).
\bigskip

\indent
The simplified approach provides a framework to study the many-body Bose gas directly in the thermodynamic limit, in terms of an equation involving a function of just 3 variables.
The method provides a promising avenue to approach singular potentials, such as the hard core.
In addition, this allows us to approach various physical questions, such as Bose-Einstein condensation, even in the intermediate density regime, away from the dilute and dense limits.

\begin{acknowledgements}
  We thank two anonymous referees for many helpful comments.
  E.H.L. thanks the Institute for Advanced study for its  hospitality.
  U.S.~National Science Foundation grants DMS-1764254 (E.A.C.),  DMS-1802170 (I.J.) are gratefully acknowledged.
\end{acknowledgements}

\bibliographystyle{apsrev4-2}
\bibliography{bibliography}

\end{document}